\input amstex
\documentstyle{amsppt}

\hsize=4.75in
\vsize=8in

\def\wick#1#2{:\!\varphi^{#1}(#2)\!:}
\def\multiwick#1#2#3#4{:\!\varphi^{#1}(#2)\cdots\varphi^{#3}(#4)\!:}
\def\multiwicktensor#1#2{:\!\varphi^{#1}\otimes\cdots\otimes\varphi^{#2}\!:}
\def\WF{\operatorname{WF}}
\def\supp{\operatorname{supp}}
\def\range{\operatorname{range}}
\def\Ad{\operatorname{Ad}}

\def\A{\Cal A}
\def\C{\Cal C}
\def\D{\Cal D}
\def\J{\Cal J}
\def\K{\Cal K}
\def\M{\Cal M}
\def\O{\Cal O}
\def\S{\Cal S}
\def\Na{\Bbb N}
\def\Re{\Bbb R}

\def\ustuff#1{\underline{#1}}
%
%
%
%
%
%
\rightheadtext{Renormalizability of $\varphi^4$}
\leftheadtext{Brunetti and Fredenhagen}
\topmatter
\title
Interacting Quantum Fields in Curved Space: Renormalizability of $\varphi^4$
\endtitle
\author
R. Brunetti$^{(1,2)}$ and K. Fredenhagen$^{(1)}$
\endauthor
\affil
$^{(1)}$II Institut f\"ur Theoretische Physik\\
Hamburg Universit\"at\\
149, Luruper Chaussee\\
D-22761 Hamburg, Germany\\
\null\\
$^{(2)}$Dip. di Scienze Fisiche\\
Universit\`a di Napoli ``Federico II''\\
Pad.19, Mostra d'Oltremare\\
I-80125, Napoli, Italy
\endaffil
\abstract
We present a perturbative construction of the $\varphi^4$ 
model on a smooth globally hyperbolic curved space-time. 
Our method relies on an adaptation of the Epstein and Glaser 
method of renormalization to curved space-times using 
techniques from microlocal analysis.
\endabstract

\footnote {To be published on the proceedings of the conference ``Operator 
Algebras and Quantum Field Theory'', held at 
Accademia Nazionale dei Lincei, Rome, Italy, July 1996.}

\endtopmatter
\document

\heading
1. Introduction
\endheading
Renormalization led to a well defined perturbation expansion of quantum 
field theory whose lowest order terms are in excellent agreement with 
experimental particle physics [1]. 
First, in the late $40$'s, quantum electrodynamic was renormalized
by the method of Schwinger, Feynman, Tomonaga and Dyson, leading to truly 
remarkable predictions, e.g., on the magnetic moment of the electron.
In the seventies, the renormalization program was extended to nonabelian 
gauge theories by the works of Faddeev-Popov, 't Hooft, Becchi-Rouet-Stora 
and others, and led to the present standard model of elementary particle 
physics. Attempts to include also gravity in the renormalization program 
failed; more recent proposals look for theories of a different kind like 
string theory which is hoped to describe all known forces.

Because of the large difference between the Planck scale ($10^{-33}$cm)
and scales relevant for the present standard model ($10^{-5}-10^{-17}$cm)
a reasonable approximation should be to consider gravity as a classical
background field and therefore investigate quantum field theory on curved 
space-times. This Ansatz already led to interesting results, the most famous 
being the Hawking radiation of black-holes [2]. But a look through the 
literature (see, e.g., [3]) shows that predominantly free field 
theories were treated on curved backgrounds. 
To our knowledge, e.g., there is no serious attempt to
discuss the influence of interaction on the Hawking effect.
Most of the papers on interacting 
quantum field theory on curved spacetime deal with the Euclidean case and 
discuss the renormalization of certain diagrams. 
There seems to exist only one attempt to a general proof of
renormalizability for $\lambda\varphi^4$, that given by Bunch [4]. However, 
also his attempt is 
confined to the rather special case of real analytic space-times which can be 
analytically continued to the Euclidean situation.
It is interesting to note that the main technical tool of this paper is 
a kind of local Fourier transformation which is a particular case of the 
powerful techniques that we use in this paper.

The situation is then tricky for general (smooth)  space-times with the
(physical) Lorentzian signature. Here more or less {\sl nothing} was done.

Why is the problem of renormalization so difficult on curved spacetime?
The main problem is absence of translation invariance. So there is no notion 
of a vacuum, which is a central object in most treatments of quantum field 
theory; the spectrum condition (positivity of the energy operator), 
responsible for deep theorems like the Spin-Statistics Theorem, cannot be 
formulated. There is no 
general connection between the Riemannian and Lorentzian field theory, and 
the meaning of the functional integral for quantum field theory on curved 
spacetime is unclear. On the more technical
side, a momentum space description is not possible, so the BPHZ method of 
renormalization [5] is not directly applicable. Also the popular method of 
dimensional renormalization seems to be restricted to the Euclidean 
situation [6].

On the other hand, physically motivated by the Equivalence Principle, 
a quick look at the possible ultraviolet divergences 
indicates that they are of the same nature as in flat space, so no obstruction
for renormalization on curved spacetime is visible. 
Despite of the interest in its own right, renormalization on curved spacetime 
might also trigger a conceptual revisitation of 
renormalization theory on flat space in the light of the principle of locality.

We sketch on this paper only the main ideas. The complete proofs should be 
found in [15].

\heading
2. The Epstein and Glaser Method
\endheading

A direct application of BPHZ or dimensional renormalization seems not to be 
possible for curved spacetime with Lorentzian signature. 
But there exists another general method developed by Epstein and Glaser [7] 
(also [8]) on the basis of ideas generally 
attributed to Bogoliubov, St\"uckelberg and their collaborators 
(see [9] and references therein). 
This method is local in spirit and is therefore our favorite candidate 
for renormalization on curved spacetime. A closer inspection shows 
that also in this method translation invariance plays an important role, both 
conceptually and technically, and it will require a lot of work to replace
translation invariance by other structures. 
In the past there has been an attempt on Minkowski 
space for quantum electrodynamics with external time independent 
electro-magnetic fields done by Dosch and M\"uller [10]. 
This use of the Hadamard parametrix of the Dirac operator is already 
much in the spirit of a local formulation of perturbation theory; by the
assumption of time independence of the external fields, however, translation 
invariance w.r.t. time still plays a crucial role in their approach.
At this point one might get the impression that a combination of techniques 
from their paper and that of Bunch (see above) will provide a useful
purely local perturbation theory. As a matter of fact, it will turn out that 
techniques from {\it microlocal analysis} [11] are ideally suited to carry 
through the program.

Let us describe the general strategy on the example of the $\varphi^4$ theory
on a $d=4$ dimensional globally hyperbolic space-time $(\M,g)$. 
We start from a quasi-free 
state $\omega$ of a free massive field $\varphi$, satisfying the Klein-Gordon 
equation of motion
$$
(\square_{g}+m^2)\varphi=0
$$
where $\square_{g}$ is the d'Alembertian operator w.r.t. the metric $g$
and where $\omega$ is supposed to satisfy the Hadamard condition [12]
$$
\omega(\varphi(x)\varphi(y))=\frac{u}{\sigma}+v\log\sigma+w
$$
with $\sigma(x,y)$ denoting the square of the geodesic distance between $x$ 
and $y$ and $u,v,w$ being smooth functions where $u,v$ are determined only by 
geometry. The commutator of the field is
$$
[\varphi(x),\varphi(y)]=iE(x,y)
$$
where $E(x,y)=E_{ret}(x,y)-E_{adv}(x,y)$, and  $E_{ret}$, resp. $E_{adv}$ are 
retarded resp. advanced Green functions which are uniquely defined on 
globally hyperbolic space-times.

As it is known one defines the ``S-matrix'' by a formal power series in the
``coupling constant''' which in the Epstein and Glaser scheme is a fixed 
test-function on space-time
$$
S(\lambda)\equiv\ \text{\bf 1}+\sum_{n=1}^{\infty}\frac{i^n}{n!}\int\ 
T_{n}(x_{1},\dots,x_{n})
\lambda(x_1)\cdots \lambda(x_n)\ d\mu_{1}\cdots d\mu_n
$$
where $\lambda\in\D(\M)$ and $d\mu$ is the natural invariant volume measure on 
$\M$ w.r.t. the fixed Lorentzian metric $g$. We remark that this definition 
is purely local thanks to the introduction of the space-time 
``coupling constant'' $\lambda$. Eventually, this test function should be sent 
to a fixed value over all space-time, but this amounts to treat the infrared 
nature of the theory to which this paper is not addressed.

In the Epstein and Glaser scheme the natural objects to use for
constructing the theory are the time-ordered products
$$
T^{k_{1},\dots,k_{n}}_{n}(x_{1},\dots, x_{n}),
\qquad k_{i}\le 4,\qquad n\in\Na
$$
which are operator-valued distributions on the GNS-Hilbert space induced by 
$\omega$. $T_{n}^{k}$ is interpreted as the time-ordered product of the 
Wick's monomials $\wick{k_{1}}{x_{1}},\dots,\wick{k_{n}}{x_{n}}$. It is
characterized by the following properties;

\roster
\item"(P1)"$\ T_{1}^{k}(x)=\wick{k}{x}$,
\item""\null
\item"(P2)"$\ T^{k_{1},\dots,k_{n}}_{n}(x_{1},\dots, x_{n})$, 
is symmetric under permutations of indices.
\endroster

Crucial is the following causality property;
\roster
\item"(P3)" If none of the points $x_1, \dots, x_l$ ($1\le l\le n$) 
lies in the past 
of the points $x_{l+1},\dots,x_{n}$, then the time-ordered product factorizes,
$$T^{k_{1},\dots,k_{n}}_{n}(x_{1},\dots, x_{n})=
T^{k_{1},\dots,k_{l}}_{l}(x_{1},\dots, x_{l})\ 
T^{k_{l+1},\dots,k_{n}}_{n-l}(x_{l+1},\dots, x_{n}).$$
\endroster

\heading 
3. Wick's Polynomials and Theorem $0$
\endheading

In the Epstein and Glaser 
scheme one requires in addition translation covariance and proves 
then that the $T_{n}$'s have an expansion into Wick's products
\roster
\item"(P4)" For any integer $n$ it holds
$$
\split T^{k_{1},\dots,k_{n}}_{n}(x_{1},\dots,&x_{n})\\
&=\sum t_{n}^{l_{1},\dots, l_{n}}(x_{1},\dots,x_{n})
\multiwick{k_{1}-l_{1}}{x_{1}}{k_{n}-l_{n}}{x_{n}}
\endsplit 
$$
\endroster
where now, the $t_{n}$'s are translation invariant numerical distributions. 
It is crucial for the program that the Wick polynomials are operator-valued 
distributions and that they can be multiplied with translation invariant 
numerical distributions. This is the content of Epstein and Glaser Theorem $0$.

On curved spacetime the first step to do is to prove the existence of Wick 
products as operator-valued distributions (even 
this was, to our knowledge, not done previous to our work [13]). Our 
construction relies on the finding of Radzikowski [14] that the wavefront set
\footnote{The wavefront set of a distribution $f$ on $\M$ 
is defined as $\WF(f)=\{(x,k)\in T^{\ast}\M\setminus\{0\}|
\phi\in\D(\M), \phi(x)\neq 0, \text{cone $C\ni k \Rightarrow$
$\widehat{\phi f}$ does not decay rapidly in $C$}\}$. Hence, 
it is a closed conic set in $T^{\ast}\M\setminus\{0\}$.}
of the two-point function of a Hadamard state is
$$
\WF(\omega_2)=\{(x,k;x^{\prime},-k^{\prime})\in T^{\ast}\M^{2}\setminus\{0\}
\ |\ (x,k)\sim (x^{\prime},k^{\prime}),\ k\in{\overline{V}_{+}}\}
$$
where the equivalence relation $\sim$ means that there exists a light-like 
geodesic from $x$ to $x^{\prime}$ such that 
$k$ is coparallel to the tangent vector to the 
geodesic and $k^\prime$ is its parallel transport from $x_1$ to $x_2$. 
The proof then uses H\"ormander's Theorem [11] that distributions can be 
pointwise multiplied provided 
the convex combinations of their wave front sets do not meet elements of the 
zero section. In our case the convexity of the forward light-cone is crucial.


The time-ordered two-point function $E_F$ (Feynman propagator) arising from 
$\omega_2$ is given by
$$
iE_F (x_1,x_2)=\omega_2 (x_1,x_2)+ E_{ret} (x_1,x_2).
$$
It has wave front set as
$$
\split
&\WF(E_F)=\\
&\ \{(x,k;x^\prime,-k^\prime)\in T^{\ast}M^2\setminus\{0\}\ |\  
(x,k)\sim (x^\prime,k^\prime), x\neq x^\prime, k\in \overline{V}_{\pm}\ 
\text{if}\  x\in \J_{\pm}(x^\prime)\}\\
&\quad\quad\quad\quad\quad\quad\cup\{(x,k;x,-k), x\in\M, k\in T^{\ast}_{x}\M
\setminus\{0\}\}
\endsplit
$$
where $\J_{\pm}(x^\prime)$ are the future/past, respectively, of $x^\prime$.

The next step amounts to replace the EG axiom of translation invariance by 
something else. We therefore assume, as an Ansatz, the expansion (P4), but 
need some restriction on the distributions $t_{n}$ which replaces the notion of
translation invariance.

We impose, instead of translation invariance, a condition on the wavefront set 
which should be satisfied by time-ordered functions. We require
\roster
\item"(P5)" It holds
$$\WF(t_{n})\subset\Gamma_{n}^{to}$$
where 
$$
\split
\Gamma_{n}^{to}=
\{&(x_{1},k_{1};\dots;x_{n},k_{n})\in 
T^{\ast}\M^{n}\setminus\{0\}\quad|\quad\exists\quad\text{a graph
$G$ with vertices}\\
&\text{$\{1,\dots, n\}$, and an association of lines $l$ 
from vertex $i=s(l)$ to $j=r(l)$}\\
&\text{of $G$ to future oriented lightlike geodesics $\gamma_{l}$ 
which connect $x_i$ and $x_j$}\\
&\text{and a covariantly constant covector field $k_{l}\in\overline{V}_{+}$ 
on $\gamma_l$ coparallel to} \\
&\text{${\dot{\gamma}}_l$ such that 
$k_{i}=\sum_{m:s(m)=i} k_{m}(x_{i})-\sum_{n:r(n)=i}k_{n}(x_{i})$}\}
\endsplit
$$
\endroster

This may be motivated by the fact that for non coinciding points $t_n$ can be 
expressed in terms of the usual Feynman graphs and for the set 
of coinciding points we have an infinitesimal remnant of translation 
invariance.

One then proves [15]: If $\WF(t_{n})\subset\Gamma_{n}^{to}$ then
$$
t_{n}(x_{1},\dots,x_{n})\multiwick{k_{1}}{x_{1}}{k_{n}}{x_{n}}
$$
is a well-defined operator-valued distribution (microlocal version of 
Theorem $0$ of Epstein and Glaser).

\heading
4. Algebraic formulation of the Epstein and Glaser approach.
\endheading

In the algebraic formulation of quantum field theory 
the basic object is a net of algebras
$$
\O\rightarrow\A(\O)
$$
where $\O$ is a relatively compact region in $\M$ and $\A(\O)$ is a von Neumann
algebra of observables localized in $\O$ which in typical cases is known 
to be a hyperfinite type III$_1$ factor. The basic hypotheses the net has to 
satisfy are

\proclaim{A1. Isotony}{\rm If $\O_1\subset\O_2$ then}
$$
\A(\O_1)\rightarrow\A(\O_2).
$$
\endproclaim

where $i_{\O_1,\O_2}$ is an injective unital homomorphism for which
$$
i_{\O_3,\O_2}\circ i_{\O_2,\O_1}=i_{\O_3,\O_1}\quad\quad\quad\text{if}\ 
\O_1\subset\O_2\subset\O_3 .
$$

\proclaim{A2. Locality} {\rm If $\O_1,\O_2\subset\O_3$ and $\O_1$ is 
spacelike separated from $\O_2$ then}
$$
i_{\O_3,\O_1}(\A(\O_1))\subset i_{\O_3,\O_2}(\A(\O_2))^\prime
$$
{\rm where $\A^\prime$ means the commutant}.
\endproclaim

One then proceeds to construct the C$^{\ast}$-inductive limit of the net
C$^{\ast}(\A)$. In the case of free field theories, the net construction 
can be made explicitely. In the interacting case, there exist constructions
only in the very special case of two dimensional Minkowski spacetime mainly
due to Glimm and Jaffe [20]. 

It seems to be less well-known that the Bogoliubov-St\"uckelberg method of
S-matrices as functionals of spacetime dependent sources actually directly
leads to a definition of local nets for interacting theories (unfortunately, 
at present, only in perturbation theory). Namely, let 
$\ustuff{g},\ustuff{h}$ be finite families of test functions on $\M$,
coupled to the various elements of the Borchers class of the free field, and 
consider the relative S-matrices
$$
V(\ustuff{g},\ustuff{h})= S(\ustuff{g})^{-1} S(\ustuff{g}+\ustuff{h}).
$$

From (P3) one finds the causality relation

\proclaim{Causality} {\rm $V(\ustuff{g},\ustuff{h}_1 +\ustuff{h}_2)=
V(\ustuff{g},\ustuff{h}_1)V(\ustuff{g},\ustuff{h}_2)$ whenever there 
are no future directed causal curves from 
$\supp \ustuff{h}_1$ to $\supp \ustuff{h}_2$}.
\endproclaim

The physical interpretation of this property is that given an 
interaction described by $\ustuff{g}$ the time evolution operator in the 
interaction picture w.r.t. any additional interaction 
has the usual factorization property. One of the main corollaries to this
condition is that it implies locality. Indeed, if $\supp \ustuff{h}_1$
is spacelike separated from $\supp \ustuff{h}_2$ 
then this is equivalent to say that no causal curves connect the two and hence
$$
V(\ustuff{g}, \ustuff{h}_1 + \ustuff{h}_2)= 
V(\ustuff{g}, \ustuff{h}_1) V(\ustuff{g}, \ustuff{h}_2)=
V(\ustuff{g}, \ustuff{h}_2) V(\ustuff{g}, \ustuff{h}_1).
$$

The local algebras of the interacting theory are now defined by
$$
\A_{\ustuff{g}}(\O)\equiv\{ V(\ustuff{g},\ustuff{h}),\ \supp 
\ustuff{h}\subset\O\}^{\prime\prime},\quad\quad 
\ustuff{g}\in\D(\O).
$$

We want to extend the definition of interacting nets $\A_{\ustuff{g}}$ to
$\C^{\infty}$-functions $\ustuff{g}$ with not necessarily compact support,
 e.g., for $\ustuff{g}=constant$. It is gratifying that this can be done
without any infrared problem.

Let $\O\subset\M$ be open and relatively compact. We define the restriction
of the net $\A_{\ustuff{g}}$ to $\O$ to be isomorphic to the net
$\A_{\ustuff{g}\phi}$ where $\phi\in\D(\M)$ with $\phi\equiv 1$ on 
a neighbourhood of $\J_{-}(\O)\cap\J_{+}(\O)$. Since the net is, up to 
isomorphy, uniquely defined by its restriction to relatively compact
open subsets of $\M$, we only have to show that 
$\A_{\ustuff{g}\phi}\restriction_{\O}$ does not depend on the choice of $\phi$.
Indeed, let $\phi^\prime\in\D(\M)$ with $\phi^\prime\equiv 1$ on a 
neighbourhood of $\J_{-}(\O)\cap\J_{+}(\O)$. Then there exist 
$\phi_{\pm}\in\D(\M)$ with $\phi^\prime=\phi+\phi_{-}+\phi_{+}$ such that 
$\J_{+}(\supp\phi_{+})\cap\J_{-}(\O)=\emptyset$ and 
$\J_{-}(\supp\phi_{-})\cap\J_{+}(\O)=\emptyset$. Let $\ustuff{h}\in\D(\O)$.
Then, by the definition of $V$, 
$$
\split
V(\ustuff{g}\phi^\prime,\ustuff{h})&=
V(\ustuff{g}(\phi+\phi_{-}),\ustuff{g}\phi_{+})^{-1}
V(\ustuff{g}(\phi+\phi_{-}),\ustuff{g}\phi_{+}+\ustuff{h})\\
&=V(\ustuff{g}(\phi+\phi_{-}),\ustuff{h})
\endsplit
$$
where the last equality follows from causality. Hence the operator 
$V(\ustuff{g}\phi^\prime,\ustuff{h})$ does not depend on the interaction 
in the future of $\supp\ustuff{h}$. It depends, however, on the interaction
in the past of $\supp\ustuff{h}$,
$$
\split
V(\ustuff{g}(\phi+\phi_{-}),\ustuff{h})&=
V(\ustuff{g}\phi,\ustuff{g}\phi_{-})^{-1}
V(\ustuff{g}\phi,\ustuff{g}\phi_{-}+\ustuff{h})\\
&=\Ad V(\ustuff{g}\phi,\ustuff{g}\phi_{-})^{-1}(V(\ustuff{g}\phi,\ustuff{h}))
\endsplit
$$
where again we used the definition of $V$ and causality. But the dependence 
is through a unitary transformation which does not depend on 
$\ustuff{h}\in\D(\O)$. Hence the nets $\A_{\ustuff{g}\phi}\restriction_{\O}$ 
and $\A_{\ustuff{g}\phi^\prime}\restriction_{\O}$ are unitarily
equivalent.
  
The interacting fields can be defined through the Bogoliubov formula
$$
\varphi_{\ustuff{g}}(\ustuff{h})=
\frac{d}{d\lambda} V(\ustuff{g},\lambda\ustuff{h})\restriction_{\lambda=0}.
$$

They may be considered as operators which are affiliated to the local 
algebras $\A_{\ustuff{g}}(\O)$. By the uniqueness above their local properties
do not depend on the behaviour of $\ustuff{g}$ outside $\O$.
In particular, one may expect that the wave front sets of their n-point
functions for a generic class of states can be determined locally.

\heading
5. Inductive Construction up to the Diagonal
\endheading

After these preparations we can mimick the argument of Epstein and Glaser 
(see also [16]) to construct $T_{n}$ 
on $\M^{n}\setminus\Delta_n$, where $\Delta_n$ is the total diagonal 
in $\M^n$, provided $T_{l}$ has been constructed for all $l< n$ and 
satisfies the causality condition (P3).

Let $J$ be the set of all $\emptyset\neq I\subsetneq\{1,\dots,n\}$. Let 
$\C_{I}=\{(x_{1},\dots,x_{n})\in\M^{n}, x_{i}\notin\J_{-}(x_{j}), i\in I,
j\in I^{c}\}$. On a globally hyperbolic space-time 
$$
\cup_{I}\C_{I}=\M^{n}\setminus\Delta_n .
$$
In fact, if $x_{i}\neq x_{j}$ for some $i\neq j$, the points $x_i$ and $x_j$ 
can be separated by a Cauchy surface $\S$, containing none of the points
$x_k$ $k=1,\dots, n$, hence $I=\{k, x_k\in\J_{+}(\S)\}\in\J$, and 
$(x_1,\dots,x_n)\in\C_{I}$.

We set on $\C_{I}$
$$
T_{I}^{k_{1},\dots,k_{n}}(x_{1},\dots,x_{n})=
T_{|I|}^{k_{i}, i\in I}(x_{i}, i\in I)
\ T_{|I^{c}|}^{k_{j},j\in I^{c}}(x_{j}, j\in I^{c}).
$$
According to the induction hypothesis and the microlocal Theorem~$0$, 
this is a well-defined operator-valued distribution on $\D(\C_{I})$. 
Different $\C_I$'s may overlap 
but one can show [15] that, due to the causality (P3) hypothesis valid for the 
lower order terms, for any $x\in\C_{I_{1}}\cap\C_{I_{2}}$ we have
$T_{I_{1}}(x)=T_{I_{2}}(x)$.

Let now $\{f_{I}\}_{I\in J}$ be a smooth partition of unity of 
$\M^{n}\setminus\Delta_n$ subordinate to $\{\C_{I}\}_{I\in J}$. Then we define
$$
{^{0}T}_{n}=\sum_{I\in J} f_{I}\ T_{I}
$$
as an operator-valued distribution on $\M^{n}\setminus\Delta_n$.

We convince ourselves that ${^{0}T}_{n}$ is independent of the choice of the 
partition of unity and symmetric under permutations of the arguments: Namely, 
let $\{f_{I}^{\prime}\}_{I\in J}$ be another partition of unity. Let 
$x\in\M^{n}\setminus\Delta_n$, and let $\K=\{I\in J, x\in\C_{I}\}$. Then there 
exists a neighbourhood $V$ of $x$ such that $V\subset\cap_{I\in\K}\C_{I}$, 
and $\supp f_{I}$ and $\supp f^{\prime}_{I}$ do not meet $V$ for all
$I\notin\K$. Then
$$
\sum_{I\in J}(f_{I}-f^{\prime}_{I})\ T_{I}\!\restriction_{V}=
\sum_{I\in\K}(f_{I}-f^{\prime}_{I})\ T_{I}\!\restriction_{V}.
$$
But on $V$, $T_{I}$ is independent of the choice of $I\in\K$. Since
$\sum_{I\in\K}f_{I}=\sum_{I\in\K}f^{\prime}_{I}=1$ on $V$, we arrive at the 
conclusion. To prove symmetry we just observe that the permuted distribution
${^{0}T}_{n}^{\pi}(x_{1}\dots,x_{n})={^{0}T}_{n}(x_{\pi(1)},\dots, 
x_{\pi(n)})$ has the expansion
$$
{^{0}T}_{n}^{\pi}=\sum_{I\in J}f_{I}^{\pi}T_{I}^{\pi}=
\sum_{I\in J}f^{\pi}_{\pi(I)}\ T_{\pi(I)}^{\pi}
$$
where we used the fact that the set $J$ is invariant under permutations, but 
$T_{\pi(I)}^{\pi}=T_{I}$ and $\{f_{\pi(I)}^{\pi}\}_{I\in J}$ is a partition 
of unity subordinate to $\{\C_{I}\}_{I\in J}$, so symmetry follows from the 
previous result on the independence of ${^{0}T}_{n}$ on the choice of the 
partition of unity.

\heading
6. The Microlocal Scaling Degree and the Extension of Distributions
\endheading

We now want to extend $^{0}T_{n}$ to the whole $\D(\M^n)$. For this purpose we 
use the expansion of $^{0}T_{n}$ into Wick polynomials
$$
{^{0}T}_{n}^{k_{1},\dots,k_{n}}=\sum\ {^{0}t}_{n}^{l_{1},\dots,l_{n}}\
\multiwicktensor{k_{1}-l_{1}}{k_{n}-l_{n}}
$$
where ${^{0}t}_{n}\in\D^{\prime}(\M^n\setminus\Delta_n)$ and 
$\WF(\chi\ {^{0}t}_{n})\subset\Gamma_{n}^{to}$ for any choice of the $l_i$ and 
any smooth function $\chi\in \C^{\infty}(\M^{n})$ such that 
$\supp\chi\subset\M^{n}\setminus\Delta_n$.

The extension to the diagonal of the tensor products of Wick's monomials 
proceeds as in our last paper [13]. Everything is therefore reduced to
the extension of the numerical distributions ${^{0}t}_{n}$ which is 
performed in two steps. 
First ${^{0}t}_{n}$ is extended by continuity to the subspace of 
test-functions which vanish on $\Delta_n$ up to a certain order, and then a 
general test-function is projected into this subspace. It is this last step 
which corresponds to the method of counterterms in the classical procedure of
perturbative renormalization. The extension of 
${^{0}t}_{n}$ by continuity requires some topology on test-function space. 
The seminorms used by Epstein and Glaser in their paper are quite 
complicated, and their generalization to curved space-times appears to be 
rather involved. 
We found it preferable therefore to apply a different method
already introduced by Steinmann [17], namely the concept of the scaling 
degree at a point of a time-ordered distribution. Its generalization to 
curved spacetime is very similar to the concept of the scaling limit as 
introduced by Haag-Narnhofer-Stein [18] and further developed by Fredenhagen 
and Haag [19].
Actually, what is really needed to implement correctly the inductive
procedure is a more general concept than the scaling degree at a
point. The requirement that the renormalized time-ordered 
distribution $t_n$ should have wave front set in $\Gamma_{n}^{to}$ 
drives us to consider a concept of scaling degree w.r.t. the diagonal 
$\Delta_n$. This is in order to get more ``uniformity'' compared to the 
pointwise case. This uniformity may be seen as a kind
of translation invariance over the diagonal. For simplicity, we first deal 
with the pointwise case and afterwards we comment on the necessary changes 
for the general one.

Let then $\M$ be a smooth manifold of dimension $d$ and $x\in\M$. 
Choose a diffeomorphism 
$\alpha$ from some convex bounded neighbourhood $V$ 
of the origin in $T_{x}\M$ onto some neighbourhood $U$ of $x$
such that $\alpha(x,0)=x$ and $d\alpha_{0}$ is the natural identification of 
the tangent space on the origin of $T_{x}\M$ with $T_{x}\M$ itself (one 
may take the exponential map for definiteness). 

Let $f\in\D^{\prime}(U)$. We define the scaled distribution 
$f_{\lambda}\in\D^\prime(U)$ by
$$
f_{\lambda}\circ\alpha(\xi)=f\circ\alpha(\lambda\xi)\quad\quad \xi\in V, 
1\ge\lambda\ge 0.
$$
Note that $f_{\lambda}$ is well defined on $U=\alpha(V)$ since by assumption 
$\lambda V\subset V$ for $0<\lambda\le 1$. 
In case $f\in\D^{\prime}(U\setminus\{x\})$ we use the above definition with 
$\xi\ne 0$ and obtain $f_{\lambda}\in\D^{\prime}(U\setminus\{x\})$.

We say that $f$ has scaling degree $\omega\in\Re$
at $x$ if $\omega$ is the smallest number such that 
$\forall\omega^{\prime}>\omega$
$$
\lim_{\lambda\downarrow 0}\lambda^{\omega^{\prime}}
f_{\lambda}(\phi)=0
$$
in the sense of distributions.
For our analysis we need a somewhat stronger version of the scaling degree 
which controls also the wavefront sets of the distributions 
$f_{\lambda}$.

\proclaim{1. Definition}
$f\in\D^{\prime}(U)$ has at $x$ the {\rm microlocal scaling degree} 
$\omega$ w.r.t. a closed cone 
$\Gamma_{x}\subset T^{\ast}_{x}U\setminus\{0\}$ if
\roster
\item"($i$)"  there exists a closed conic set 
$\Gamma\subset T^{\ast}U\setminus\{0\}$ with 
$\Gamma\cap T_{x}^{\ast}U\subset \Gamma_{x}$ such that 
$\WF(f)_{\lambda}\subset\Gamma$ for sufficiently small $\lambda$.
\item"($ii$)" $\omega$ is the smallest number such that for all 
$\omega^\prime >\omega$
$$
\lim_{\lambda\downarrow 0}\lambda^{\omega^{\prime}}
f_{\lambda}=0
$$
in the sense of the H\"ormander pseudotopology on $\D^{\prime}_{\Gamma}(U)$.
\endroster
\endproclaim

We recall that by the H\"ormander pseudotopology it is meant the 
following [11];
given a sequence of distributions 
$u_{i}\in\D^{\prime}_{\Gamma}(\M)\equiv\{v\in\D(\M)\ |\ \WF(v)\subset\Gamma\}$,
we say that the sequence converges to $u$ in the 
sense of H\"ormander pseudotopology in $\D^{\prime}_{\Gamma}(\M)$ 
whenever the following two properties hold true:
\roster
\item"(a)" $u_{i}\to u$ weakly,
\item"(b)" for any properly supported pseudodifferential operator $A$ such
that $\WF(A)\cap\Gamma=\emptyset$, we have that $Au_i \to Au$ in the sense of
$\C^{\infty}(\M)$
\endroster
where $\WF(A)$ is defined by the projection in $T^{\ast}\M$ of the wave front
set of the Schwartz kernel associated to $A$.

For $f\in\D^{\prime}(U\setminus\{x\})$ we cannot directly define the 
microlocal scaling degree. We require instead that for all 
$\chi\in\C^{\infty}(U)$ with $x\notin\supp\chi$ the sequence $\chi f_\lambda$
(considered as a distribution on $U$) satisfies the two conditions of the 
Definition above.

The microlocal scaling degree ($\mu$sd) has nice properties. 
For $f_1$ and $f_2$ with 
$\mu$sd $\omega_1$ and resp. $\omega_2$ at $x$, w.r.t. $\Gamma^1_x$ resp. 
$\Gamma^2_x$ such that
$\{0\}\notin(\Gamma^{1}_x +\Gamma^{2}_x)$, the wave front sets of 
$f_{1,\lambda}$ and $f_{2,\lambda}$ for sufficiently small $\lambda$ satisfy
the condition 
$(\WF(f_{1,\lambda})\oplus\WF(f_{2,\lambda}))\cap\{0\}=\emptyset$, hence their
product exists [11] and, because of the sequential continuity of the products
in the H\"ormander pseudotopology, have $\mu$sd $\omega\le \omega_1 +\omega_2$
w.r.t. 
$\Gamma_x = \Gamma^{1}_x\cup\Gamma^{2}_x\cup(\Gamma^{1}_x +\Gamma^{2}_x)$.

We now want to show how to extend the distribution 
$f\in\D^{\prime}(U\setminus\{x\})$ to all space. We first deal with the 
extension problem using the scaling degree. The next section contains the 
proof of the extension problem w.r.t. the $\mu$sd. 
There are two possible cases:
when the scaling degree $\omega\ge d$ or otherwise $\omega< d$. We first study
the second case. 

\proclaim{2. Theorem} If $f_0\in\D^{\prime}(U\setminus\{x\})$ has scaling 
degree $\omega< d$ at $x$ then there exists a unique $f\in\D^{\prime}(U)$ 
with $f(\phi)=f_0(\phi)$, $\phi\in\D(U\setminus\{x\})$ and the same scaling 
degree.
\endproclaim
\demo{Proof} Let $\vartheta\in\D(U)$ with
$\vartheta\equiv 1$ on a neighbourhood of $x$. We define for $0<\lambda <1$
$$
\vartheta_{\lambda^{-1}}(x)=\cases
\vartheta(\alpha(\lambda^{-1}\alpha^{-1}(x))), & x\in\alpha(\lambda V)\\
0,&\text{else}.
\endcases
$$
Then $1-\vartheta_{\lambda^{-1}}\in\C^{\infty}(U)$ with 
$x\notin\supp(1-\vartheta_{\lambda^{-1}})$. We want to define $f$ as the limit
$$
f\equiv\lim_{n\to\infty}f_0(1-\vartheta_{2^n}).
$$
Let $\phi\in\D(U)$. Then $f_0(1-\vartheta_{2^n})(\phi)$ is a Cauchy sequence.
Namely, for $n>m$
$$
\split
f_0(1-\vartheta_{2^n})(\phi)-f_0(1-\vartheta_{2^m})(\phi)
&=f_0(\vartheta_{2^m}-\vartheta_{2^n})(\phi)\\
&=\sum_{j=m}^{n-1}(\phi f_0)(\vartheta_{2^j}-\vartheta_{2^{(j+1)}})\\
&=\sum_{j=m}^{n-1}(\phi f_0)_{2^{-j}}(\vartheta-\vartheta_{2}) 2^{-jd}
\endsplit
$$
where we used the definition of the scaled distribution as well as an 
identification of densities and functions by the use of a measure $d\mu$
with $\alpha^{\ast}d\mu=d\xi$, the latter denoting the Lebesgue measure on
$T_{x}\M$.

As a smooth function $\phi$ has scaling degree equal to $0$. Hence 
$\forall\omega^\prime >\omega$ there is a constant $c$ such that 
$$
|(\phi f_0)_{2^{-j}}(\vartheta-\vartheta_{2})|\le c\ 2^{j\omega^\prime}.
$$

We insert this estimate for $\omega<\omega^\prime <d$ and obtain the desired 
result. 

It remains to prove that $f$ has the same scaling degree as $f_0$. Let
$\omega<\omega^\prime <d$. Let $\phi\in\D(U)$. Then
$$
(1-\vartheta_{2^n})\phi_{\lambda^{-1}}=0
$$
if $n<n_\lambda$ for some $n_\lambda\in\Na$ with 
$2^{-n_{\lambda}}\lambda^{-1}\to_{\lambda\to 0}\text{const.}\ne 0$. Hence
$$
\split
\lambda^{\omega^\prime}f_{\lambda}(\phi)&=\lambda^{(\omega^\prime-d)}
f(\phi_{\lambda^{-1}})\\
&=\lim_{n\to\infty}\lambda^{(\omega^\prime-d)}
(f_0 (1-\vartheta_{2^n}))(\phi_{\lambda^{-1}})\\
&=\sum_{n\ge n_{\lambda}}\lambda^{(\omega^\prime-d)}
f_0((\vartheta_{2^n}-\vartheta_{2^(n+1)})\phi_{\lambda^{-1}})\\
&=\sum_{n\ge n_{\lambda}}\lambda^{(\omega^\prime-d)} 2^{-nd}
(f_0)_{2^{-n}}((\vartheta-\vartheta_{2})\phi_{\lambda^{-1}2^{-n}}).
\endsplit
$$
The test-function has all Schwartz norms uniformly bounded  in 
$\lambda$ and $n\ge n_\lambda$. Hence
$$
|(f_0)_{2^{-n}}((\vartheta-\vartheta_{2})\phi_{\lambda^{-1}2^{-n}})|\le 
c\ 2^{n\omega^{\prime\prime}}
$$
for some $\omega<\omega^{\prime\prime}<\omega^\prime$. But then
$$
\split
|\lambda^{\omega^\prime}f_{\lambda}(\phi)|&\le
\sum_{n\ge n_{\lambda}}\lambda^{(\omega^\prime-d)} 
2^{-n(d-\omega^{\prime\prime})}\\
&\le\lambda^{(\omega^\prime-d)} 
\frac{2^{-n(d-\omega^{\prime\prime})}}{1-2^{-(d-\omega^{\prime\prime})}}\\
&\le\lambda^{(\omega^\prime-\omega^{\prime\prime})} 
\frac{(\lambda^{-1}2^{-n_{\lambda}})^{(d-\omega^{\prime\prime})}}
{1-2^{-(d-\omega^{\prime\prime})}}\to 0.  
\endsplit
$$
 The uniqueness is obvious since any other extension differs by a derivative of the delta
function based at the point $x$ which has scaling degree 
$\ge d$. $\quad\blacksquare$
\enddemo

For the case when $f_0$ has scaling degree $\omega\ge d$ we deal here only
with the preliminary step of extension on a subspace 
of test-functions $\D_{\delta}(U)\subset\D(U)$
whose derivatives at the point $x$ vanish up to order $\delta=[\omega]-d$. 

\proclaim{3. Theorem} Let $f_0\in\D^{\prime}(U\setminus\{x\})$ have scaling
degree $\omega\ge d$. Then the sequence $f_0((1-\vartheta_{2^n})\phi)$ with
$\phi\in\D_{\delta}(U)$, $\delta=[\omega]-d$, 
converges and the limit defines a unique distribution
$\overline{f}(\phi)\equiv\lim_{n\to\infty}f_0((1-\vartheta_{2^n})\phi)$ over 
$\D_{\delta}(U)$.
\endproclaim
\demo{Proof} (Sketch) The proof goes similar to the one of Theorem 2. 
The only change is that now $\phi$ has scaling degree $\le -\delta-1$ 
and the estimate would change as follows
$$
|(f_0 \phi)(\vartheta_{2^m}-\vartheta_{2^n})|\le c\ 
\sum_{j=m}^{n-1}2^{j(\omega^\prime-\delta-1-d)}
$$
hence, choosing $\omega^\prime$ such that the exponent is negative we get the
convergence. That $\overline{f}$ is a distribution follows from the 
Banach-Steinhaus 
Theorem applied to $\D_{\delta}(U)$ which is a closed subset of $\D(U)$. 
 $\quad\blacksquare$
\enddemo

In the application of this procedure to the n-th order of perturbation 
theory we want to scale only in the difference 
variables. On a curved space-time this might be done in the following way. We 
choose a map $\alpha:\ T\M\to\M$ such that $\alpha(x,0)=x$ and 
$d\alpha(x,\ \cdot\ ){\!\restriction_{0}}=\text{id}$ (for instance, the 
exponential map). We then define 
$\alpha_{n}:\ T\M^{n}\!\restriction_{\Delta}\to\M^n$ by
$$
\alpha_{n}(x,\xi_{1},\dots,\xi_{n})
=(\alpha(x,\xi_{1}),\dots,\alpha(x,\xi_{n})).
$$

We restrict $\alpha_n$ to the following sub-bundle which is isomorphic to the 
normal bundle of $\Delta_n$
$$
N\Delta_n =\{(x,\xi_{1},\dots,\xi_{n})\in T\M^{n}\!\restriction_{\Delta_n}, 
\sum\xi_{i}=0\}.
$$

For a sufficiently small neighbourhood of the zero section in $N{\Delta_n}$
$\alpha_n$ restricted to it 
is a diffeomorphism onto some neighbourhood of $\Delta_n$. We now
express our time-ordered function as a distribution on $N{\Delta_n}$ and do 
the scaling w.r.t. the variables $\xi$. There is however a complication with 
this procedure; namely, in the inductive construction, the coordinates so
obtained do not factorize, hence it is not obvious how the microlocal 
scaling degree
of lower order terms determines the microlocal scaling degree at higher order.
Much easier is the behaviour of the total scaling degree, w.r.t. all variables.
Here it is easy to see that the scaling degree of the factors determines the 
scaling degree for tensor and pointwise products.
We therefore prove a Lemma which states that the condition on the wave 
front set
of time-ordered distributions implies that they can be restricted to the
submanifolds
$$
\M_{x}=\{\alpha_{n}(x,\xi_{1},\dots,\xi_{n}), \sum\xi_{i}=0\}.
$$
Again the result on the continuity of restrictions in the H\"ormander 
pseudotopology gives us the desired information on the microlocal 
scaling degree.

\proclaim{4. Lemma}
Let $\WF({^{0}t}_{n})\subset\Gamma_{n}^{to}$. 
Then $\forall x\in\M$ there exists 
$\phi\in\D(\M^n)$ with $\phi(x,\dots,x)\neq 0$, such that 
$\phi {^{0}t}_{n}$ can be restricted to $\M_{x}$.
\endproclaim
\demo{Proof} It suffices to show that the wavefront set of ${^{0}t}_n$ 
in a 
neighbourhood of $(x,\dots,x)$ does not intersect the conormal bundle of the 
submanifold $\M_{x}\subset\M^n$. But at the point $(x,\dots,x)$ the elements 
$(x,k_{1},\dots,k_{n})$
of the wavefront set, with 
$(k_{1},\dots, k_{n})\neq 0$, satisfy $\sum k_{i}=0$, hence the equation
$$
\sum \langle k_{i},\xi_{i}\rangle=0
$$
which characterizes a point in the conormal bundle of $\M_{x}$, 
cannot hold for all $(\xi_{1},\dots,\xi_{n})$ with $\sum \xi_{i}=0$.

Now the wavefront set intersected with the conormal bundle of $\M_{x}$ is a 
closed conic subset of $T^{\ast}\M^n$ which does not contain 
$T^{\ast}_{(x,\dots,x)}\M^n$, hence does not contain 
also a conic neighbourhood of 
$T^{\ast}_{(x,\dots,x)}\M^n$, but such a neighbourhood always contains a set 
$T^{\ast}U$ where $U$ is a neighbourhood of $(x,\dots,x)$. If we choose $\phi$ 
with support in $U$ we arrive at the desired conclusion. $\quad\blacksquare$
\enddemo

We now want to impose some condition on the smoothness of the above 
construction w.r.t. $x$ which serves as a substitute for translation
invariance. Let $t$ be a distribution from $\D^{\prime}(\M^n)$ whose wave
front set is orthogonal to the tangent bundle of the diagonal, i.e., 
$$
\langle\xi,k\rangle=0, \quad x\in\Delta_n,\quad \xi\in T_{x}\Delta_n, 
\quad (x,k)\in\WF(t).
$$

We set 
$$
\tilde{t}_{\lambda}(x,\alpha_n (x,\xi))=t(\alpha_n (x,\lambda\xi))
\quad\quad (x,\xi)\in T\M^n\!\restriction_{\Delta_n}.
$$
$\tilde{t}_{\lambda}\equiv (1\otimes t)_\lambda$ is a distribution on a 
neighbourhood $\tilde{U}$ of the diagonal 
$\Delta_{n+1}$  in $\M^{n+1}$. We say that $t$ has $\mu$sd $\omega$ at
$\Delta_n$ if there is a closed conic set 
$\tilde{\Gamma}\subset T^{\ast}\tilde{U}$ with
$$
\langle\xi,k\rangle=0, \quad\forall \xi\in T_{x}\Delta_{n+1},\quad 
(x,k)\in\tilde{\Gamma}
$$
such that
\roster
\item"($i$)" $\WF(\tilde{t}_\lambda)\subset\tilde{\Gamma}$, 
for all sufficiently small $\lambda$,
\item"($ii$)" $\omega$ is the smallest number such that for all 
$\omega^\prime >\omega$
$$
\lambda^{\omega^\prime}\tilde{t}_{\lambda}\to 0
$$
in the sense of $\D_{\tilde{\Gamma}}^{\prime}(\tilde{U})$.
\endroster

We then restrict $\tilde{t}_{\lambda}$ to the submanifold 
$\cup_{x\in\M}(\{x\}\times\M_{x})\subset\M^{n+1}$. This is possible by 
the argument in the proof of Lemma 4. Note that the submanifold 
$\cup_{x\in\M}(\{x\}\times\M_{x})$ 
might be identified with the open set $U\equiv\cup_{x\in\M}\M_{x}\subset\M^n$.

Let $t_\lambda$ denote the restriction of $\tilde{t}_\lambda$ to $\D(U)$. By
the sequential continuity of the restriction operator we find that 
$$
\lambda^{\omega^\prime}t_\lambda\to 0\quad\quad\omega^\prime >\omega
$$
in the sense of $\D^{\prime}_{\Gamma}(U)$ as $\lambda\to 0$.

We can now calculate the microlocal scaling degree of 
time-ordered distributions, given by pointwise and/or tensor products
ot lower order time-ordered distributions.

For the Feynman propagator we find that the $\mu$sd is $d-2$ at every 
point of the diagonal, w.r.t. the $\WF(\Delta_{F})$ on flat space.

If we assume that $t^{l}_{n}$ has $\mu$sd $\omega_{n}^l$ 
at the diagonal in $\M^n$ w.r.t. $\Gamma_{n}^{to}$ then
$$
t_{I,L}^{l_{1},l_{2}}=(t^{l_{1}}_{|I|}\otimes t^{l_{2}}_{|I^{c}|})
\prod_{L}\Delta_{F}(x_{i},x_{j})
$$
has $\mu$sd at the diagonal equal to 
$\omega_{|I|}^{l_{1}}+\omega_{|I^{c}|}^{l_{2}}+(d-2)|L|$ w.r.t. 
$\Gamma_{n}^{to}$.
Hence the $\mu$sd of ${^{0}t}_{n}^{l}$ is determined by the $\mu$sd at 
lower orders. We now have to study whether ${^{0}t}_{n}^{l}$  can be
extended to all $\D(\M^n)$ such that the $\mu$sd is conserved.

\heading
7. Extension to the Diagonal
\endheading

In the last Section we saw that the time-ordered functions 
${^{0}t}_n\equiv {^0}t$, 
originally defined only on $\D(\M^{n}\setminus\Delta_n)$, can be extended to 
$\D(\M^n)$ or $\D_{\delta}(\M^n)$, where $\delta=[\omega]-(n-1)d$, 
whenever the $\mu$sd $\omega$, which is computed in terms of the $\mu$sd
of the time-ordered functions at lower orders, satisfies either 
$\omega<(n-1)d$ or $\omega\ge (n-1)d$. Note that the presence of the term 
$(n-1)d$ is related to our choice of the relative coordinates.
We now want to remove the restriction in the second case 
by simply projecting arbitrary  test-functions onto 
$\D_{\delta}(\M^n)$. It is this last step which corresponds in other 
renormalization schemes to the subtraction of infinite counterterms.

Here we do the projection in the following way. We choose a function $w$ which 
is equal to $1$ on a neighbourhood of $\Delta_n$ and with support in 
$\range(\alpha_n)$,
where the map $\alpha_n$ of the last Section is used in order to introduce 
relative coordinates. We set
$$
(W\phi)(x_1,\dots, x_n)=
\phi(x_1,\dots,x_n)-w(x_1,\dots,x_n)
\sum_{|\beta|\le\delta}\frac{\xi^{\beta}}{\beta!}\partial^{\beta}_{\xi}
(\phi\circ\alpha_{n})(x,\xi=0)
$$
with $\alpha_{n}(x,\xi)=(\alpha(x,\xi_1),\dots,\alpha(x,\xi_n))$, 
$\xi=(\xi_1,\dots,\xi_n)$ and
the usual multi-index notations for $\xi^\beta$ and $\partial_{\xi}^\beta$ and
define, following the Theorems 2 and 3, 
$t(\phi)\equiv{\overline{{^0}t}}(W\phi)$.

If we would apply ${^{0}t}$ to the single terms in the definition of $W\phi$
the first term would correspond to the divergence and the second one to the 
counterterm. This can be made explicit by choosing a sequence of smooth 
functions ${^{k}t}$ on $\M^n$ which converges to $t$ in the 
sense of distributions. Then
$$
{t}=\lim_{k\to\infty}\left( {^{k}t}-\sum_{\beta}\langle 
{^{k}t}, w\frac{\xi^{\beta}}{\beta!}\rangle\ 
\delta^{(\beta)}(\xi)\right).
$$

Let us now reconsider the extension problem. If 
${^0}t\in\D^{\prime}(\M^n\setminus\Delta_n)$ we say that it has at 
$\Delta_n$ $\mu$sd $\omega$ w.r.t. $\Gamma^{to}_n$ if, for all 
$\chi\in\C^{\infty}(\M^{n+1})$ with 
$(\M\times\Delta_n)\cap\supp\chi=\emptyset$, the two following 
properties hold true
\roster
\item"($i$)" $\WF(\chi(1\otimes {^0}t)_\lambda )\subset\tilde{\Gamma}^{to}_{n}=
\{(x,0;y,k), x\in\M, (y,k)\in\Gamma^{to}_n\}$,
\item"($ii$)" $\lambda^{\omega^\prime}\chi(1\otimes{^0}t)_\lambda\to 0$ in 
$\D^{\prime}_{\tilde{\Gamma}^{to}_{n}}$.
\endroster

Choosing $\vartheta\in\C^{\infty}(\M^n)$, $\vartheta\equiv 1$ on a 
neighbourhood of $\Delta_n$ with $\supp\vartheta\cap\M_x $ a compact 
set for any $x\in\Delta_n$, we recall that the extensions of ${^0}t$ 
are obtained by
$t\equiv\lim_{n\to\infty}(1-\vartheta_{2^n}){^0}t$ whenever $\omega<(n-1)d$
or by $t\equiv\lim_{n\to\infty}(1-\vartheta_{2^n}){^0}t\circ W$ whenever 
$\omega\ge (n-1)d$, where 
$\vartheta_{2^n}(\alpha_n(x,\xi))=\vartheta(\alpha_n(x,2^n\xi))$ with
$(x,\xi)\in N\Delta_n$. 

We can prove that the sequences converge in $\D^{\prime}_{\Gamma^{to}_n}(U)$ 
and keep the same $\mu$sd at $\Delta_n$ of their respective ${^0}t$'s. We 
recall that $U\equiv\cup_{x\in\M}\M_x$. The convergence in the sense of 
distributions is given by Theorems 2 and 3, the one in the smooth sense, after 
application of a suitable pseudodifferential operator, will  
 be done in the following way. Note that we consider the proof 
only in the case when $\omega\ge (n-1)d$ the proof of the other trivially 
follows by simply choosing $W$ equal to the identity operator.

\proclaim{5. Theorem} If $W:\D(U)\to\D_{\delta}(U)$ is the restriction to $U$ 
of the map $W$ defined above, and $\delta=[\omega]-(n-1)d$, the expression
$$
{^0}t (1-\vartheta_{2^m})\circ W
$$
converges to $t$ as $m\to\infty$ in the H\"ormander 
pseudotopology for $\D_{\Gamma_{n}^{to}}^{\prime}(\M^n)$.
\endproclaim
\demo{Proof} 
Since in Section 6 we already proved the convergence in the sense of 
distributions we need only to check convergence in the smooth sense of the 
sequence of smooth functions $AW^t\ {^0}t (1-\vartheta_{2^m})$ 
for the appropriate pseudo-differential operators $A$ 
(see remark after Definition 1.).
 Note that the functions
$1-\vartheta_{2^m}$ are such that their product
with ${^0}t$ gives a distribution in $\M^n$. The argument for proving 
the convergence in the smooth sense goes similar to the proof of the 
convergence in distribution sense. 

For a pseudodifferential operator with smooth kernel the result follows from 
the convergence in the sense of distributions. By subtracting from $A$ an 
operator with smooth kernel, if necessary, we may assume that the kernel of
$A$ has support in a sufficiently small neighbourhood of the diagonal in
$\M^n\times\M^n$. Since for $\phi\in\D(\M^n\setminus\Delta_n)$ we have
$$
{^0}t(1-\vartheta_{2^m})(W\phi)={^0}t(\phi)
$$
we may restrict the consideration to a sufficiently small neighbourhood of 
$\Delta_n$.

In terms of local coordinates on $\M^n$, $A$ may be described, in a 
neighbourhood of a point on the diagonal, in terms of its symbol
$$
(Au)(x,\xi)=\int\int\sigma_{A}(x,\xi;p,k)\hat{u}(p,k)dp dk
$$
where $\sigma_{A}$ is fast decreasing in a conical neighbourhood of
$\Gamma_n^{to}\cap T^{\ast}U$. Since $\Gamma_{n}^{to}$ contains the 
conormal bundle of $\Delta_n$, the symbol $\sigma_{A}$ decays fast in a cone
around the point $p=0$,
$$
|\sigma_{A}(x,\xi;p,k)|\le\ c_{N}(1+|p|+|k|)^{-N}
$$
where the constants $c_N$ are independent of $(x,\xi)\in U$.

We now want to apply $A$ to the distribution 
${^0}t (1-\vartheta_{2^m})\circ W$. Using the Lagrange formula for the 
rest term in the Taylor expansion, we find for the action $u\to u\circ W$
the expression
$$
u\circ W=u(1-w)+Pu
$$
where the first term vanishes near the diagonal and may therefore be ignored, 
and where in local coordinates $(x,\xi)$, the Fourier transformation of the
second term is
$$
\widehat{Pu}(p,k)=\int^{1}_{0}dt \sum_{|\beta|=\delta+1} 
\widehat{uw\xi^{\beta}}(p,tk) k^{\beta} \frac{(1-t)^{\delta}}{\delta!} .
$$

Therefore we obtain for sufficiently large $m$
$$
\split
|(A {{^0}t}& (\vartheta_{2^m}-\vartheta_{2^{m+1}})\circ W)(x,\xi)|
\le\text{const.}\ 2^{-m(n-1)d}\ \sum_{\beta} \int 
|\sigma_{A}(x,\xi;p,k)|\\
&\times
|\widehat{({^0}t w\xi^\beta)_{2^{-m}}(\vartheta-\vartheta_{2})}(p,2^{-m}tk)|
|k|^{\beta}\ dp\ dk\ dt .
\endsplit
$$
Because of the assumption on the $\mu$sd of ${^0}t$ we have the estimate
$$
|\widehat{({^0}t w\xi^{\beta})_{2^{-m}}
(\vartheta-\vartheta_{2})}(p,2^{-m}tk)|
\le c_N\ 
(1+|p|+|2^{-m}tk|)^{-N} 2^{m(\omega^\prime -\delta-1)}
$$
for every closed cone which does not contain $p=0$.

Since $\sigma_A$ and 
$\widehat{({^0}t w\xi^\beta)_{2^{-m}}(\vartheta-\vartheta_2)}$ 
are polynomially bounded
we conclude that the integrand is fast decreasing and we get
$$
|(A {{^0}t_n} (\vartheta_{2^m}-\vartheta_{2^{m+1}})\circ W)(x,\xi)|
\le\text{const.}\  2^{m(\omega^\prime -\delta-1-(n-1)d)} .
$$
Since $\delta=[\omega]-(n-1)d$ then the exponent becomes equal to
$\omega^\prime -[\omega]-1$ which for a choice of a 
sufficiently small $\omega^\prime$ is negative, hence the thesis follows.
$\quad\blacksquare$

\enddemo

It remains to compute the $\mu$sd of $t_{n}^{l}\equiv t$. We first check 
the scaling degree definition and we get
$$
\split
\tilde{t}_{\lambda}(\phi)&= \tilde{t}(\phi^{\lambda}),\quad\text{where}\ 
\phi^{\lambda}(\xi)=\lambda^{-(n-1)d}\phi(\lambda^{-1}\xi)\\
&= {^{0}\tilde{t}}\left(\phi^{\lambda}-
w\sum\frac{\xi^{\beta}}{\beta!}\ \partial_{\xi}^{\beta}\phi^\lambda(0)\right)\\
&= {^{0}\tilde{t}}\left(\phi^{\lambda}-
w(\lambda^{-1}\ \cdot\ )\sum\frac{(\lambda^{-1}\xi)^{\beta}}{\beta!}
\ \partial_{\xi}^{\beta}\phi(0)\ \lambda^{-(n-1)d}\right)\\
&\quad\quad\quad\quad\quad+\sum {^{0}\tilde{t}}
\bigl((w(\lambda^{-1}\ \cdot\ )-w)(\lambda^{-1}\xi)^{\beta}\bigr)
\ \frac{\partial_{\xi}^{\beta}\phi(0)}{\beta!}\ \lambda^{-(n-1)d}\\
&= {^{0}\tilde{t}}_{\lambda}(W\phi)+
\sum\frac{\partial_{\xi}^{\beta}\phi(0)}{\beta!}\
{^{0}\tilde{t}}_{\lambda}\bigl((w-w_{\lambda})\xi^{\beta}\bigr) .
\endsplit
$$

The first term scales as assumed. The term of the sum over $\beta$ can be 
analyzed in the following way. We write
$$
w-w_{\lambda}=\int_{\lambda}^{1}\frac{d}{d\mu}w_{\mu}d\mu .
$$

We convince ourselves that the integral can be commuted with the application 
of the distribution. We find
$$
\split
{^{0}\tilde{t}}_{\lambda}\left( (\frac{d}{d\mu}w_{\mu})\xi^{\beta}\right)&=
{^{0}\tilde{t}}_{\lambda}\bigl( (\partial_{i}w)(\mu\xi)\xi^{i}\xi^{\beta}
\bigr)\\
&= {^{0}\tilde{t}}_{\lambda}\bigl( (\partial_{i}w)(\mu\xi)(\mu\xi)^{i}
(\mu\xi)^{\beta}\bigr)\mu^{-|\beta|-1}\\
&= {^{0}\tilde{t}}_{\lambda/\mu}\bigl({(\partial_{i}w)(\xi)\xi^{i}\xi^{\beta}}
\bigr)\mu^{-|\beta|-1-d(n-1)} .
\endsplit
$$

But by assumption
$$
|{^{0}\tilde{t}}_{\lambda/\mu}\left((\partial_{i}w)\xi^{i}\xi^{\beta}\right)|
\le\text{const.}\left(\frac{\lambda}{\mu}\right)^{-\omega^{\prime}}\quad\quad
\forall\omega^{\prime}>\omega,
$$
hence
$$
\split
| {^{0}\tilde{t}}_{\lambda}&\left((w-w_{\lambda})\xi^{\beta}\right)|
\le\text{const.}
\ \lambda^{-\omega^{\prime}}\int_{\lambda}^{1}d\mu
\mu^{\omega^{\prime}-|\beta|-1-(n-1)d}\\
&\null\\
&\le\text{const.}\ \lambda^{-\omega^{\prime}}\ \cdot\ 
\cases
\frac{1-\lambda^{\omega^{\prime}-|\beta|-d(n-1)}}
{\omega^{\prime}-|\beta|-d(n-1)}, 
   &\text{for $\omega^{\prime}-|\beta|-d(n-1)> 0$}\\
|\ln\lambda|,&\text{for $\omega^{\prime}-|\beta|-d(n-1)= 0$}.
\endcases
\endsplit
$$

We conclude that we obtain the same formula for the singularity degree as in 
the well-known power counting rules. Indeed, we get that $|\beta|\le
\omega -(n-1)d$, $\omega$ being the infimum over all $\omega^\prime$. 
Now, $\omega$ can be computed since for
$t^{k_1,\dots,k_n}_n$ we have $\omega =\sum_i k_i (d-2)/2$ where 
$(d-2)/2$ is the canonical dimension of the scalar field as can be seen from
its two point function on Minkowski space. Collecting the formulas 
we get, taking $k_i=4$ for all $i=1,\dots,n$, that
$|\beta|\le 2n(d-2) -(n-1)d$
hence $|\beta|\le n(d-4)+d$ which for $d=4$ does not depend on $n$ anymore.
This implies that the renormalization prescription works well since the 
number of counterterms does not grow up with the induction step. 

It remains to check the convergence in the smooth sense after application 
of a properly supported pseudodifferential operator whose ``wave front set'' 
is disjoint from $\Gamma_{n}^{to}$. We get from the formula above an identity 
for distributions as
$$
\lambda^{\omega^\prime}A\tilde{t}_{\lambda}= \lambda^{\omega^\prime} 
A ({^0}\tilde{t}_{\lambda}\circ W) + \lambda^{\omega^\prime}
\sum_{\beta}\ (-1)^{|\beta|}\  A ({^0}\tilde{t}_{\lambda}
((w-w_{\lambda})\frac{\xi^{\beta}}{\beta!})\delta^{(\beta)}(\xi)) .
$$
The first term has been already discussed and scales
as $\lambda^{\omega^\prime -\delta -1 -(n-1)d}$. For the second
one, the above proof of the convergence in H\"ormander pseudotopology can 
be redone almost word by word since the application of $A$ 
gives a smooth function. One finally finds the same scaling 
behaviour as above by looking at the behaviour of the term 
${^0}\tilde{t}_{\lambda}((w-w_{\lambda})\xi^{\beta})$ as
 $\lambda\to 0$. Hence we get convergence also in the smooth sense, 
provided the same choice of $\omega^\prime$ is done consistently with the 
discussion done until now.

\heading
8. Conclusions
\endheading
 
To summarize: The inductive procedure gives
symmetric, renormalized time ordered distributions which satisfy
the causality condition. Moreover, we have shown how these renormalized objects
satisfy the microlocal requirements in terms of wave front set and microlocal
scaling degree. We found that the criterion for renormalizability follows 
the same power counting rules as on Minkowski space. All that by purely local 
methods.

It is now important to remove the remaining ambiguity
by fixing the finite renormalization. We hope to report elsewhere on this last 
attempt [15]. At this stage several questions arise:
\roster
\item"$(a)$" Do the interacting fields satisfy the $\mu$SC of [13]?
\item"$(b)$" How can the renormalization group be treated? (see, e.g., [21])
\item"$(c)$" Is there a corresponding Euclidean formulation [22]?
\item"$(d)$" Can the construction be extended to gauge theories?
\item"$(e)$" What are the gravitational corrections to quantum field effects?
\item"$(f)$" How does the interaction modify the Hawking radiation?
\endroster

\heading
9. Acknowledgements
\endheading

Special thanks are due to Martin K\"ohler for help during the early 
stage of this project. We also thank Raymond Stora for helpful suggestions
and for communications of unpublished material on the Epstein and Glaser
method. The first author wishes to thank the members of the II Institut f\"ur
Theoretische Physik, Hamburg Universit\"at, for warm hospitality during 1996. 
His research was supported by EEC under contract n. ERBFMBICT950355 of
the Training and Mobility of Reaserchers Programme.

\heading
References
\endheading

\item{[1]} S.~Weinberg: {\it The quantum theory of fields}, vol.I-II, 
Cambridge University Press, 1995-1996.

\item{[2]} S.~Hawking: {\it Particle creation by black holes}, 
Comm. Math. Physics. {\bf 43}, (1975) p.199.

\item{[3]} R.~Wald: {\it Quantum field theory in curved spacetime and black 
hole thermodynamics}, Chicago University Press, 1994. Compare also; 
N.~D.~Birrel and P.~C.~W.~Davies: {\it Quantum fields in curved space}, 
Cambridge University Press, 1982; and S.~Fulling: {\it Aspects of quantum 
field theory in curved space-time}, Cambridge University Press, 1989.

\item{[4]} T.~S.~Bunch: {\it BPHZ Renormalization of $\lambda\phi^4$ 
field theory in curved spacetime}, Annals of Physics, {\bf 131}, (1981), p.118.

\item{[5]} C.~Itzykson and J-B.~Zuber: {\it Quantum field theory}, 
McGraw-Hill, 1980.

\item{[6]} M.~L\"uscher: {\it Dimensional regularisation in the presence of
large background fields}, Annals of Physics, {\bf 142}, 1982, p.359.

\item{[7]} H.~Epstein and V.~Glaser: {\it The role of locality in perturbation 
theory}, Ann. Inst. Henri Poincar\'e-Section A, vol. XIX, n.3, (1973), p.211.

\item{[8]}  G.~Scharf: {\it Finite quantum electrodynamics: The causal 
approach}, Springer-Verlag, 1995, 2nd edition.

\item{[9]} N.~N.~Bogoliubov and D.~V.~Shirkov: {\it Introduction to the 
theory of quantized fiels}, John Wiley and Sons, 1976, 3rd edition.

\item{[10]} H.~G.~Dosch and V.~F.~M\"uller: {\it Renormalization of quantum
electrodynamics in an arbitrary strong time independent external field}, 
Fortschritte der Physik, {\bf 23}, (1975), p.661.

\item{[11]} There are several books available on the subject, among them a 
selection (due to our uncomplete knowledge of the existing literature) 
might be;
\roster
\item"(a)" L.~H\"ormander: {\it The analysis of linear partial differential
operators}, vol. I-IV, Springer-Verlag, 1983-1986.
\item"(b)" F.~Treves: {\it Introduction to pseudodifferential and Fourier 
integral operators}, vol. I-II, Plenum Press, 1980.
\item"(c)" M.~E.~Taylor: {\it Partial differential equations}, vol. I-III, 
Springer-Verlag, 1996.
\endroster

\item{[12]} B.~Kay and R.~Wald: {\it Theorems on the uniqueness and thermal 
properties of stationary, non singular, quasifree states on spacetimes with a 
bifurcate Killing horizon}, Phys. Rep., {\bf 207}, (1991), p.49.

\item{[13]} R.~Brunetti, K.~Fredenhagen and M.~K\"ohler: {\it The microlocal
spectrum condition and the Wick's polynomials of free fields}, Comm. Math. 
Physics, {\bf 180}, (1996), p.312.

\item{[14]} M.~Radzikowski: {\it Micro-local approach to the Hadamard 
condition in quantum field theory on curved space-time}, Comm. Math. Physics, 
{\bf 179}, (1996), p.529;
and, {\it A local-to-global singularity theorem for quantum field theory on 
curved space-time}, Comm. Math. Physics, {\bf 180}, (1996), p.1.

\item{[15]} R.~Brunetti and K.~Fredenhagen: {\it Microlocal analysis and 
interacting quantum field theory}, papers in preparation.

\item{[16]} R.~Stora: {\it Differential algebras in Lagrangean field theory}, 
ETH Lectures, January-February 1993; and also, G.~Popineau and R.~Stora: 
{\it A pedagogical remark on the main theorem of perturbative renormalization 
theory}, unpublished preprint.

\item{[17]} O.~Steinman: {\it Perturbation expansions in axiomatic field 
theory}, Lect. Notes in Physics, {\bf 11}, Springer-Verlag, 1971.

\item{[18]} R.~Haag, H.~Narnhofer and U.~Stein: {\it On quantum field theory 
in gravitational background}, Comm. Math. Physics, {\bf 94}, (1984), p.219.

\item{[19]} K.~Fredenhagen and R.~Haag: {\it Generally covariant quantum 
field theory}, Comm. Math. Physics, {\bf 108}, (1987), p.91.

\item{[20]} J.~Glimm and A.~Jaffe: {\it Quantum physics. A functional 
integral point of view}, McGraw-Hill, 1981.

\item{[21]} I.~L.~Bukbinder, S.~D.~Odintsov, I.~L.~Shapiro: {\it 
Renormalization group approach to quantum field theory on curved space time},
Riv. Nuovo Cimento, {\bf 12}, 1989, n.10 p.1.

\item{[22]} W.~Junker: work in progress.

\enddocument
\bye